\documentstyle[twoside,fleqn,espcrc2]{article}

\newcommand{\ms}{\mbox{{\footnotesize{$\overline{{\rm MS}}$}}}}
\newcommand{\fiv}{\mbox{\footnotesize{$5+\overline{5}$}}\ }
\newcommand{\ten}{\mbox{\footnotesize{$10+\overline{10}$}}\ }
\def\gsim{\kern.4em\raise.3ex
\hbox{$>$\kern-.75em\lower1ex\hbox{$\sim$}}\kern.4em}
\def\lsim{\kern.4em\raise.3ex
\hbox{$<$\kern-.75em\lower1ex\hbox{$\sim$}}\kern.4em}

\newcommand{\AmS}{{\protect\the\textfont2
  A\kern-.1667em\lower.5ex\hbox{M}\kern-.125emS}}

\title{$\chi^2$ Analysis of Supersymmetric Models\thanks{Talk given by
D.M.P. at the 5th International Conference on Supersymmetries in
Physics (SUSY 97), Philadelphia, Pennsylvania, May 27-31,
1997. D.M.P. is supported by Department of Energy contract
DE--AC03--76SF00515.}}

\author{Damien M. Pierce\address{Stanford Linear Accelerator Center,
Stanford University, Stanford, California 94309, USA} and Jens
Erler\address{Institute for Particle Physics, University of
California, Santa Cruz, California 95064, USA}}

\begin{document}

\begin{abstract}
We discuss the results of a global fit to precision data in
supersymmetric models. We consider both gravity- and gauge-mediated
models. As the superpartner spectrum becomes light, the global fit to
the data typically results in larger values of $\chi^2$. We indicate
the regions of parameter space which are excluded by the data. We
discuss the additional effect of the $B(B\rightarrow X_s\gamma)$
measurement. Our analysis excludes chargino masses below $M_Z$ in the
simplest gauge-mediated model with $\mu>0$, with stronger constraints
for larger values of $\tan\beta$.
\end{abstract}

\maketitle

\section{Introduction}

Low energy measurements can serve as useful probes of higher energy
scales, because the virtual effects of heavy particles influence low
energy observables. Hence, low energy measurements can constrain
possible new physics scenarios.  The most striking example of this
effect was the ``virtual top-quark discovery''. When the top-quark
mass was first measured through direct production at the Tevatron, the
precision electroweak data had already constrained the mass with about
the same central value and uncertainty \cite{DMP:top}.

In the standard model some observables are sensitive to the square of
the top-quark mass and the logarithm of the Higgs boson mass. Hence,
both these masses can be constrained by precision data. In
supersymmetric models, some observables are sensitive to the
supersymmetric masses. Just as with $m_t$ and $M_H$, there are values
of supersymmetric parameters in conflict with measurements. We will
discuss the results of a global fit to precision data in
supersymmetric models.

Before discussing the supersymmetric case, we review the results of a
global fit within the standard model. This serves as a useful
barometer for comparison with the supersymmetric models. Also, the
supersymmetric models reduce to the standard model as the
supersymmetric mass scale becomes large. The supersymmetric
corrections decouple as $M_Z^2/M_{\rm SUSY}^2$ (or faster). The only
remnant of supersymmetry in the large $M_{\rm SUSY}$ limit is the
light Higgs boson mass (which remains a prediction of the model).

\section{The observables}

\begin{table*}[th]
\setlength{\tabcolsep}{1.5pc}
\newlength{\digitwidth} \settowidth{\digitwidth}{\rm 0}
\catcode`?=\active \def?{\kern\digitwidth}
\begin{tabular}{|l|c||c|c|}
\hline
                    & Measurement       &        Standard       &  Pull \\
                    &                   &         Model         &       \\
\hline
\hline
$M_Z$ [GeV]                 & 91.1863  $\pm$ 0.0019   &  91.1862 &\ \ 0.0\ \ \\
$\Gamma_Z$ [MeV]            &  2494.7  $\pm$ 2.6      &  2496.9   &$-$0.9 \\
$\sigma_h$ [nb]             & 41.489   $\pm$ 0.055    &  41.467   &   0.4  \\
$R_e$                       & 20.756   $\pm$ 0.029    &  20.757   &   0.0 \\
$R_\mu$                     & 20.795   $\pm$ 0.029    &  20.757   &   1.0 \\
$R_\tau$                    & 20.831   $\pm$ 0.029    &  20.802   &   0.5 \\
$A_{\rm FB}^e$              &  0.0161  $\pm$ 0.0010   &   0.0162  &   0.0 \\
$A_{\rm FB}^\mu$            &  0.0165  $\pm$ 0.0010   &   0.0162  &   0.2 \\
$A_{\rm FB}^\tau$           &  0.0204  $\pm$ 0.0010   &   0.0162  &   2.3 \\
\hline
${\cal A}_\tau(\tau)$       &  0.1401  $\pm$ 0.0067   &  0.1469   &$-$1.0 \\
${\cal A}_e(\tau)$          &  0.1382  $\pm$ 0.0076   &  0.1469   &$-$1.1 \\
$\sin^2\theta^{\rm lept}_{\rm eff} (\langle Q_{FB}\rangle)$
                            &  0.2322  $\pm$ 0.0010   &  0.2315   &   0.7  \\
\hline
$R_b$                       &  0.2177  $\pm$ 0.0011   &   0.2158  &   1.7 \\
$R_c$                       &  0.1722  $\pm$ 0.0053   &   0.1723  &   0.0 \\
$A_{\rm FB}^b$              &  0.0985  $\pm$ 0.0022   &   0.1030  &$-$2.1 \\
$A_{\rm FB}^c$              &  0.0735  $\pm$ 0.0048   &   0.0736  &   0.0 \\
${\cal A}_b$                &  0.897   $\pm$ 0.047    &   0.935   &$-$0.8  \\
${\cal A}_c$                &  0.623   $\pm$ 0.085    &   0.667   &$-$0.5 \\
\hline
$A_{LR}$                    &  0.1548  $\pm$ 0.0033   &   0.1469  &   2.4  \\
${\cal A}_\mu$              &  0.102   $\pm$ 0.034    &   0.147   &$-$1.3  \\
${\cal A}_\tau$             &  0.195   $\pm$ 0.034    &   0.147   &   1.4  \\
\hline
$Q_W$(Cs)                   &$-$72.11  $\pm$ 0.93     &$-$73.11   &   1.1   \\
$Q_W$(Tl)                   &$-$114.77 $\pm$ 3.65     &$-$116.7   &   0.5   \\
$M_W$ [GeV]                 &  80.402  $\pm$ 0.076    &   80.375  &   0.4   \\
$m_t$ [GeV]                 &  175.6   $\pm$ 5.0      &  173.0    &   0.5   \\
$\Delta\alpha_{\rm had}$    &  0.028037$\pm$ 0.000654 & 0.02797   &   0.1   \\
\hline
\end{tabular}             
\caption{Measured and best fit values of the observables in the
standard model.}
\end{table*}

In the standard model we take as inputs the muon decay constant,
$G_\mu$, the $Z$-boson mass, $M_Z$, the top quark mass, $m_t$,
the Higgs boson mass, $M_H$, the electromagnetic coupling, $\alpha$,
and the strong coupling, $\alpha_s$. The last two couplings
are taken in the \ms\ scheme at the $Z$-scale.

Given these six inputs\footnote{We need to specify the remaining
fermion masses as well, but the predictions for the observables we
consider are not very sensitive to these inputs.} we have predictions
for all other observables in the standard model. We consider the list
of observables below, and find the values of the five
inputs\footnote{Because of the small error, we take $G_\mu$ as a fixed
input.} which minimize the total $\chi^2$. In this way we find the best
fit values of the input parameters and the standard model predictions
for all the observables.

The observables we include in our $\chi^2$ are

\begin{itemize}
\item Line-shape and lepton asymmetries\footnote{We include the
correlations among these nine observables.}. These are the $Z$-mass,
the $Z$-width, the peak hadronic cross-section, the ratio of the
hadronic width to the leptonic widths, and the leptonic forward-backward
asymmetries: $M_Z$, $\Gamma_Z$, $\sigma_{\rm had}$,
$R_{e,\mu,\tau}$, $A^{FB}_{e,\mu,\tau}$.

\item $\tau$ polarization. The $\tau$ decay analysis yields
measurements of the $\tau$ and $e$ left-right asymmetries: ${\cal
A}_{\tau,e}(\tau)$.

\item Light quark charge asymmetry, $\langle Q_{FB}\rangle$, which
yields a measurement of $\sin^2\!\theta^{\rm lept}_{\rm eff}$.

\item $b$ and $c$ quark results. These are the ratios of the heavy
quark widths to the total hadronic width, and the heavy quark
forward-backward asymmetries (polarized at the SLC): $R_{b,c}$, 
$A^{FB}_{b,c}$, ${\cal A}_{b,c}$.

\item Leptonic left-right asymmetries (total and forward-backward): 
$A_{LR}$, ${\cal A}_{\mu,\tau}$.

\item Atomic parity violation weak charges: $Q_W$(Cs) and $Q_W$(Tl).

\item $W$-boson mass, top-quark mass, and the light quark contribution
to $\alpha$: $M_W$, $m_t$ and $\Delta\alpha_{\rm had}$.

\end{itemize}

There are 26 observables included here. With 5 parameters in the fit,
we are left with 21 degrees of freedom.

\section{The standard model fit}

In Table 1 we list the results of the standard model fit. We list the
measured values with the errors\footnote{The data is current as of
spring 1997.} \cite{DMP:LEPEWWG,DMP:SLD,DMP:APV,DMP:mt,DMP:mw,DMP:EJ},
the standard model predictions, and the pull, which is defined to be the
difference between the measured value and the standard model prediction, 
divided by the error. If we sum the squares of the pulls we obtain the
$\chi^2$ of the best fit, which in this case is 29.57. With 21 degrees
of freedom, this corresponds to a probability of 10.1\%. This means
that if the standard model does describe the data, in a random variate
(a set of measured values with central values distributed randomly in
a Gaussian fashion) the probability that the $\chi^2$ is greater than
29.57 is 10.1\%. This 10.1\% probability may sound low, but we consider it
to be reasonable. Remember that the ``best'' we could hope for is 50\%
(higher than that would suggest that the data fits the predictions
``too well''). In the days when $R_b$ was 3.5$\sigma$ off, the
goodness of the fit was much less than 1\%.

The best fit values of some inputs are
$$m_t = 173\pm5\ {\rm GeV}\ ,$$
$$\alpha_s = 0.122 \pm 0.003\ ,$$
$$M_H = 93 ^{+104}_{-57}\ {\rm GeV}\ .$$
We show contours of constant $\chi^2$ in the $m_t-M_H$ plane in Fig.~1.
The contours correspond to 68\% and 90\% confidence. We see that
$M_H$ is constrained to be less than 400 GeV at 95\% CL.

\begin{figure}[htb]
\vbox{\kern4cm\includegraphics{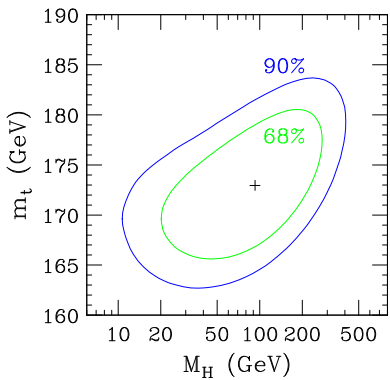}}
\caption{The standard model 68 and 90\% CL contours in the $m_t$,
$M_H$ plane.}
\end{figure}

\section{The supersymmetric analysis}

There are two approaches to $\chi^2$ analyses in supersymmetric models
which one might consider. In the first approach one tries to repair
the discrepancies seen between the data and the standard model
\cite{DMP:prec.mssm}. For example, the standard model prediction for
$R_b$ used to be 3.5$\sigma$ off. One could find regions of
supersymmetric parameter space where this discrepancy was repaired
\cite{DMP:Rb}. Now, however, there are no large ($3\sigma$)
discrepancies, and the largest deviations cannot be repaired by
supersymmetry. This approach is no longer useful, since the
supersymmetric corrections cannot reduce the $\chi^2$ significantly,
and one has to pay the price of the smaller number of degrees of
freedom.

In the other approach, one notices that there are significant regions
of parameter space where the supersymmetric corrections make the fit
worse. Consider the plot\footnote{This plot corresponds to the
``minimal supergravity'' model with the universal soft parameters set
to $M_{\rm SUSY}$, $\tan\beta=2$ and $\mu>0$. Some region of the curve
has superpartners lighter than the current limits.} of $\chi^2$ vs.
$M_{\rm SUSY}$ in Fig.~2. At large $M_{\rm SUSY}$ the $\chi^2$
approaches the standard model value (with the light Higgs mass
determined as a function of the supersymmetric parameters). At smaller
$M_{\rm SUSY}$, the $\chi^2$ typically rises; the unsuppressed
supersymmetric radiative corrections can result in a terrible
fit. Here we focus on this approach, elucidating the regions of
parameter space where the $\chi^2$ is so large that those points can
be ruled out.

\begin{figure}[htb]
\vbox{\kern4cm\includegraphics{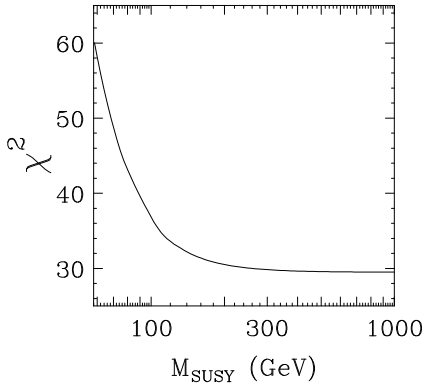}}
\caption{$\chi^2$ vs. $M_{\rm SUSY}$.}
\end{figure}

Notice there are two competing effects which will continue to
determine the utility of this approach. As the data becomes more
precise, the smaller errors will lead to larger values of $\chi^2$
(assuming the discrepancies are real). At the same time, as more data
is accumulated, the limits on the supersymmetric mass spectrum are
increased. One must then consider larger values of the supersymmetric
masses, and this leads to smaller values of $\chi^2$.

\subsection{Three supersymmetric models}

We will show the results of the $\chi^2$ analysis in three
supersymmetric models. The ``minimal supergravity'' model
\cite{DMP:msugra} is perhaps the most commonly considered high scale
model. Here we take as boundary conditions at the GUT scale (the scale
where $g_1=g_2$) a universal scalar mass $M_0$, a universal gaugino
mass $M_{1/2}$, and a universal trilinear scalar coupling $A_0$. We
also take as an input the ratio of vacuum expectation values of the
two Higgs doublets, $\tan\beta\equiv v_2/v_1$, and the sign of the
$\mu$-term.

Consider two assumptions which are necessary in order to arrive at the
``minimal supergravity'' model. First, one must assume a flat K\"ahler
metric.  Second, one assumes the universality of the scalar masses is
maintained from the Planck scale to the GUT scale. Both of these
assumptions seem artificial and unrealistic, especially the
latter. However, they are useful for simplicity and economy, and in
particular they guarantee that the model is free of potentially
disastrous FCNC problems. Also, the running of the parameters between
the Planck scale and the GUT scale is model dependent, so it makes
sense that the minimal model assumes no running.

Models with gauge-mediated (GM) supersymmetry breaking \cite{DMP:gmsb}
comprise a class which is automatically free of FCNC problems. Here,
the supersymmetry breaking is communicated from the hidden sector to
the visible sector through the interactions of the gauge fields and
the messenger fields. In the minimal model the superpotential contains
a singlet which acquires both a vev $X$ and an $F$-term $F_X$. The
singlet is coupled to the messenger fields, which are in a vector-like
representation under the standard model gauge group. The
supersymmetric spectrum is proportional to $\Lambda\equiv F_X/X$, with
dependence on the messenger field representation. We consider two
models, a model with a \fiv messenger sector and a model with a \ten
messenger sector. The boundary conditions for the soft masses are
applied at the messenger mass scale, $M$. Again, we take $\tan\beta$
and the sign of $\mu$ as inputs.

In both the supergravity and gauge-mediated models we impose radiative
electroweak symmetry breaking \cite{DMP:ewsb}. Starting with a common
positive Higgs boson mass-squared at the GUT scale or the messenger
scale, we evolve the Higgs masses down to the weak-scale using the
renormalization group equations (RGE's) \cite{DMP:rge's}. Because of
the large top-quark Yukawa coupling, the mass-squared of the Higgs
which couples to the top, $m_{H_2}^2$, is driven negative in the
vicinity of the electroweak scale. This signals the breaking of
electroweak symmetry, and requiring this to occur allows us to solve
for the heavy Higgs boson mass and the Higgsino mass as a function of
the input parameters:
\begin{eqnarray}
m_A^2 &=& {1\over\cos2\beta}\left(m_{H_2}^2-m_{H_1}^2\right)-M_Z^2 \\
\mu^2 &=& {1\over2} \biggl[ \tan2\beta\left( m_{H_2}^2\tan\beta -
m_{H_1}^2 \cot\beta \right) \\
&& \qquad\qquad\qquad\qquad\qquad- M_Z^2\biggr]\nonumber
\end{eqnarray}
In the gauge-mediated models we implicitly assume that whatever
mechanism is responsible for the generation of the $B$ and $\mu$ terms
does not give rise to contributions to the soft scalar masses.

\subsection{The determination of $\chi^2$}\label{DMP:proc}

The overview of the $\chi^2$ analysis is as follows:

\begin{enumerate}

\item Pick starting values for $(M_Z,\ m_t,\ \alpha,\ \alpha_s)$.

\item Pick a random point in supersymmetric parameter space.

\item For fixed $(M_Z,\ m_t,\ \alpha,\ \alpha_s)$
solve the supersymmetry model by iteration.

Here we have two-sided boundary conditions. We know the gauge and
Yukawa couplings and $\tan\beta$ at the weak scale and the soft
parameters at the high scale. We include full one-loop corrections in
the evaluation of the gauge and Yukawa couplings, and in the Higgs
sector (both the light Higgs boson mass and electroweak symmetry
breaking).

\item Compute $\chi^2$. Here we include the full one-loop
supersymmetric corrections to every observable, with two caveats:

\begin{itemize}
\item We use the oblique approximation for the evaluation of atomic
parity violation weak charges
\item The SUSY box diagrams are neglected in $Z$-pole observables
\end{itemize}

\item Minimize $\chi^2$ with respect to $(M_Z, m_t, \alpha,
\alpha_s)$ for the fixed set of supersymmetric corrections.

\item If not converged, go to step 3.

\item Apply current limits on the superpartner and Higgs boson mass
spectrum from direct searches. If this fails, disregard this point.

\end{enumerate}

We include the current mass limits from CDF, D0 \cite{DMP:cdfd0}, and
LEP II \cite{DMP:lep2}.  LEP II has produced new limits on the
chargino mass, the slepton masses, the Higgs boson masses, and the
light top-squark mass. Because the spectrum in high scale models is
correlated, the CDF and D0 gluino and squark mass limits are typically
irrelevant. For example, in the \ten gauge-mediated model, after
imposing the limits on the chargino, Higgs bosons, and sleptons, we
find that the squark and gluino masses are larger than 260 GeV. The
direct search limits are 230 and 180 GeV, respectively
\cite{DMP:cdfd0}.

\subsection{The oblique approximation}

We can give a concise description of the overall magnitude and
relevance of the supersymmetric corrections by considering the oblique
approximation. Most of the precision observables involve gauge boson
exchange, and hence they receive universal (i.e. process and
flavor-independent) corrections from the gauge boson self-energies.
With $M_Z$, $G_\mu$, and $\alpha$ as inputs, there are three
independent linear combinations of gauge-boson self-energies which
appear in physical observables in the lowest order of a derivative
expansion. In some cases the full corrections are dominated by the
oblique corrections. This might be expected, since the oblique
corrections include contributions from every non-singlet superpartner,
so they are enhanced by the number of generations and/or the number of
colors. The non-oblique corrections (the fermion wave-function,
vertex, and box corrections) arise from a limited set of diagrams,
since the loops are constrained by the external fermion quantum
numbers.

\begin{figure}[htb]
\vbox{\kern5.2cm\includegraphics{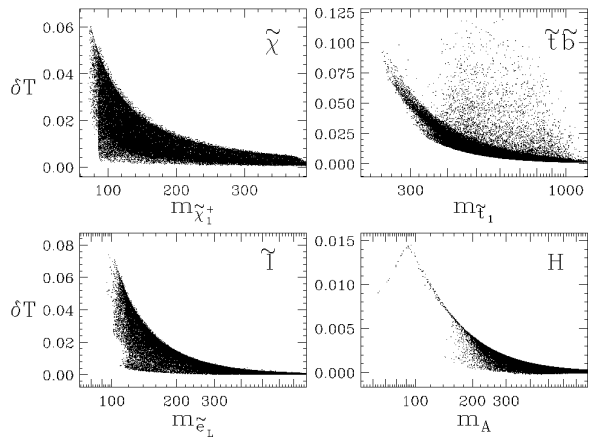}}
\caption{The contributions to $T$ from the chargino/neutralino,
stop/sbottom, slepton and Higgs sectors, in Figs.~(a), (b), (c) and
(d), respectively, vs. the representative masses. The masses are
in units of GeV.}
\end{figure}

\begin{figure*}[htb]
\vbox{\kern3.7cm\includegraphics{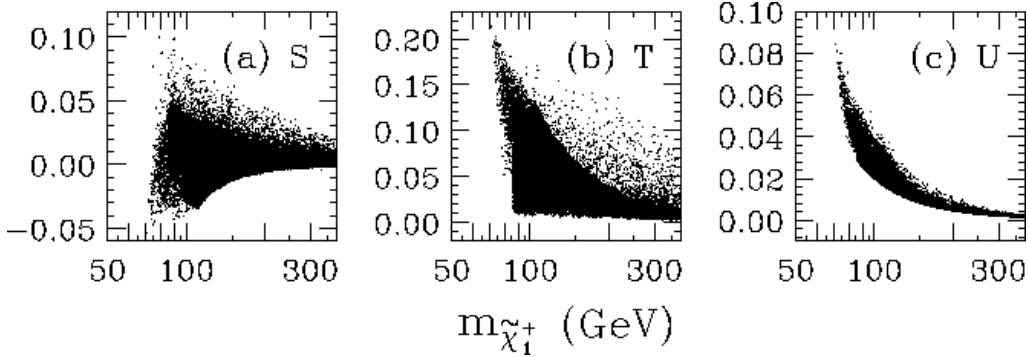}}
\caption{The supersymmetric contributions to $S$, $T$ and $U$ 
in the supergravity model.}
\end{figure*}

We parametrize the oblique corrections by $S$, $T$ and $U$
\cite{DMP:stu}. These are given by the expressions \cite{DMP:stu form}
\begin{eqnarray}
S&=& \biggl[\cos^2\theta_W\left(F_{ZZ} - F_{\gamma\gamma}\right)\\
&&\qquad- {\cos\theta_W\over\sin\theta_W}\cos2\theta_W F_{\gamma Z}\biggr]
\times {4\sin^2\theta_W\over\alpha}\nonumber\\
T&=& \biggl[{\Pi_{WW}(0)\over M_W^2} - {\Pi_{ZZ}(0)\over M_Z^2}\\
&&\qquad\qquad  -2{\sin\theta_W\over\cos\theta_W}{\Pi_{\gamma Z}(0)
\over M_Z^2}\biggr]
\times {1\over\alpha}\nonumber\\
U&=&\biggl[F_{WW} - \cos^2\theta_WF_{ZZ} -
\sin^2\theta_WF_{\gamma\gamma} \\
&&\qquad\qquad -\sin2\theta_WF_{\gamma Z}\biggr]
\times{4\sin^2\theta_W\over\alpha}\nonumber
\end{eqnarray}
where $F_{ij}=(\Pi_{ij}(M_j^2)-\Pi_{ij}(0))/M_j^2$ (except
$F_{\gamma\gamma}=\Pi_{\gamma\gamma}(M_Z^2)/M_Z^2$).

The most important of these parameters is $T$. In Fig.~3 we show the
contributions to $T$ from the various supersymmetric sectors in the
supergravity model. Each point in the scatter plots is a best fit at
the randomly chosen point in supersymmetric parameter space, as
described in Section~\ref{DMP:proc}. Fig.~3(a) shows the
chargino/neutralino contribution vs. the light chargino mass, (b)
shows the stop/sbottom contribution vs. the heavy stop mass, (c) shows
the slepton contribution vs. the left-handed selectron mass, and (d)
shows the supersymmetric Higgs boson contribution vs. the CP-odd Higgs
boson mass. We note that all the sectors give positive contributions
to $T$ (also to $U$). We see that the first three sectors contribute
with the same order of magnitude (at most about 0.06, 0.12, and 0.07,
respectively). The Higgs sector and the contributions of the first two
generation squarks each contribute at most about 0.015.

If we add all these contributions together, we obtain the plots in
Fig.~4. Here we show the total supersymmetric contributions to $S$,
$T$ and $U$ vs. the light chargino mass. The cancellation between the
slepton and chargino sectors results in typically smaller values for
$S$. The SUSY contributions to $S$, $T$ and $U$ are in the ranges
($-0.05,0.1$), (0,0.2), and (0,0.09). For chargino masses above 300
GeV the decoupling results in suppressed contributions.

\begin{figure*}[htb]
\vbox{\kern3.7cm\includegraphics{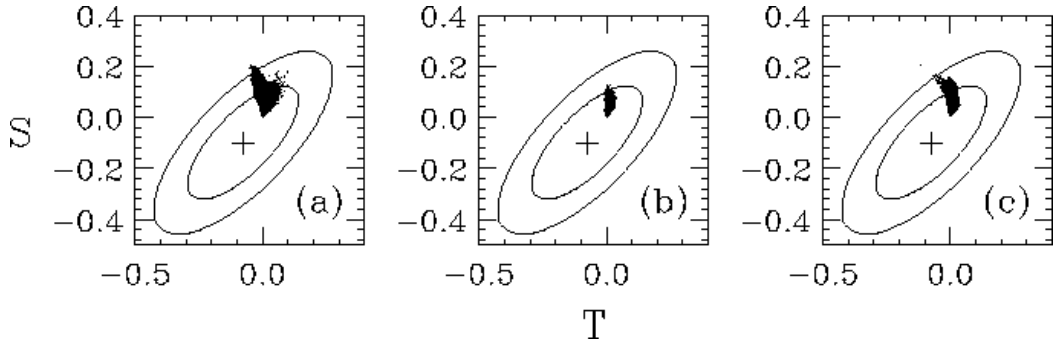}}
\caption{The supersymmetric contributions to $S$ and $T$ on the
68\% and 95\% standard model contours. The supergravity
model, the \fiv GM model, and the \ten GM model scatter plots are
shown in Figs. (a), (b) and (c), respectively.}
\end{figure*}

We illustrate the relevance of these corrections in Figs.~5. Here, in
the $T$, $S$ plane, we show the 68\% and 95\% CL contours found by
varying $M_Z$, $m_t$, $\alpha$, $\alpha_s$, and $U$,
with $M_H$ fixed to its best fit value, in the standard model.  On top
of these contours we show scatter plots of the supersymmetric
contributions to $S$ and $T$. In Fig.~5(a) the minimal supergravity
scatter plot is shown, in 5(b) the \fiv gauge-mediated model scatter
plot is shown, and the \ten GM model results are shown in
5(c). The point (0,0) is a part of each supersymmetric scatter plot,
since this corresponds to the decoupled region of parameter space,
where $M_{\rm SUSY}$ is large. As we decrease $M_{\rm SUSY}$, the
scatter moves up and to the right or left. One can get a feeling for
the overall magnitude of the supersymmetric corrections here. The
direct limits on the supersymmetric mass spectrum have significantly
constrained the magnitude of the supersymmetric corrections. For
example, in the \fiv gauge-mediated model the contributions to $S$ and
$T$ fall almost entirely inside the 68\% contour.

\subsection{Full one-loop analysis}

\begin{figure}[htb]
\vbox{\kern2cm\includegraphics{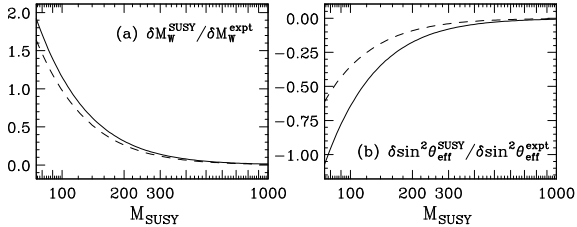}}
\caption{The supersymmetric corrections to (a) $M_W$ and (b)
$\sin^2\theta_{\rm eff}^{\rm lept}$ vs. $M_{\rm SUSY} = M_0 = M_{1/2}
= A_0$, with $\tan\beta=2$ and $\mu>0$.  The solid line shows the full
one-loop correction, and the dashed line indicates the oblique
approximation.}
\end{figure}

To rule out points in parameter space we need to consider the full
one-loop supersymmetric corrections, not just the oblique
corrections. The oblique approximation works well for some
observables, but poorly for others. We illustrate this in Fig.~6.  In
Fig.~6(a) we show the $W$-boson mass vs. $M_{\rm SUSY}$ in the oblique
approximation and the full one-loop result, and they are seen to be
in good agreement (here we work in the supergravity model, define
$M_{\rm SUSY} = M_0 = M_{1/2} = A_0$, and set $\tan\beta=2$ and
$\mu>0$).  We show the correction in units of ``pull'' (that is, we
divide the correction by the experimental error). In Fig.~6(b) we show
the full and oblique $\sin^2\theta^{\rm lept}_{\rm eff}$. Clearly the
non-oblique corrections are significant in this case.

In the full one-loop supersymmetric analysis we have included an
external constraint on the strong coupling, $\alpha_s=0.118 \pm0.003$,
which we obtained by combining all except the $Z$-lineshape data
\cite{DMP:alphas}.

\begin{figure*}[htb]
\vbox{\kern10cm\includegraphics{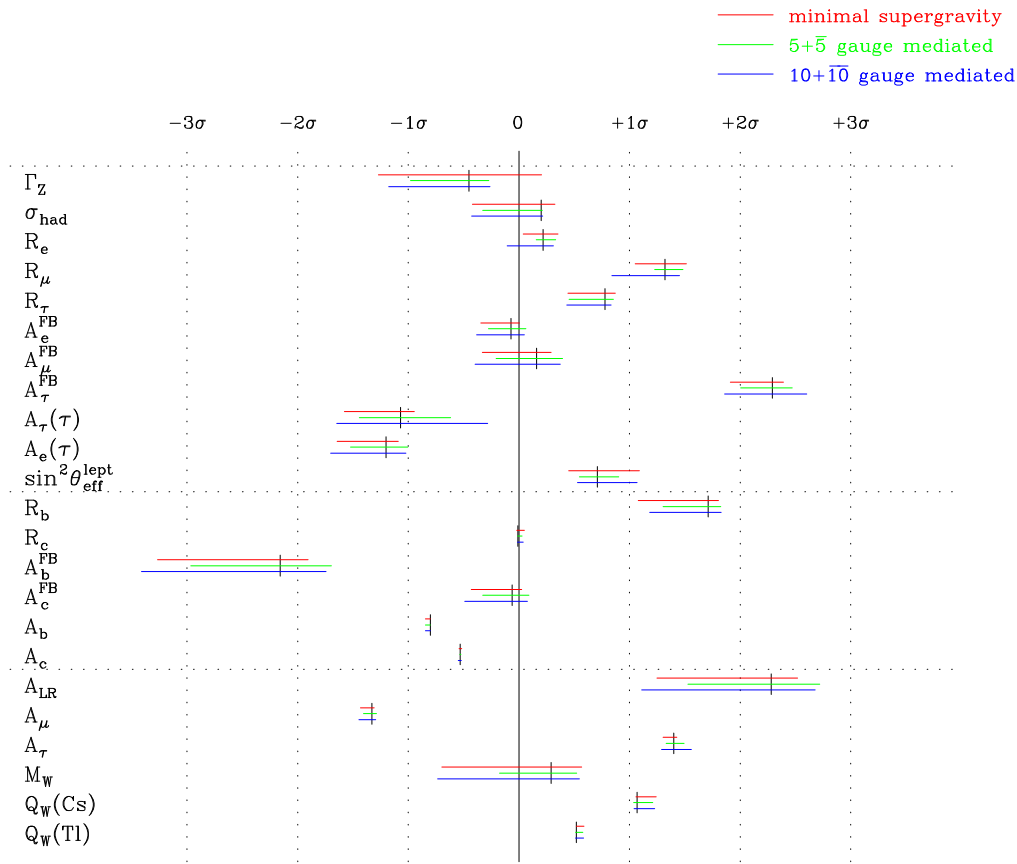}}
\caption{The maximal extent of the supersymmetric corrections to each
observable, in units of pull. The horizontal lines correspond to, from
top to bottom, the supergravity model, the \fiv GM model, and the \ten
GM model. The standard model pull is indicated by the small vertical
lines.}
\end{figure*}

The extent of the complete one-loop corrections for each observable is
shown in Fig.~7, where the horizontal lines indicate the range of
values of each observable in the entire supersymmetric parameter
space. The top line corresponds to the supergravity model, the middle
line to the \fiv GM model, and the bottom line to the \ten GM
model. The small vertical line shows the value of the pull in the
standard model. 

Looking at this plot the general impression is that the supersymmetric
corrections are unable to provide a significantly more satisfactory
description of the data than the standard model.  Some of the
observables are seen to be irrelevant in the fit (i.e. $R_c,\ {\cal
A}_c,\ {\cal A}_b,\ {\cal A}_\mu,\ {\cal A}_\tau,\ Q_W({\rm Cs})$ and
$Q_W({\rm Tl})$; the supersymmetric corrections to these observables
are small relative to the experimental uncertainty). Of the
observables with sizable corrections, some can reduce the SM
discrepancies, ($R_b$, $A_{LR}$), while others can increase the SM
discrepancies (${\cal A}_{e,\tau}(\tau)$, $A^{FB}_b$).  There are
interesting correlations among the corrections to different
observables. For example, the positive corrections to $R_b$ (which
reduce the SM discrepancy) are accompanied by positive corrections to
$A^{FB}_b$ (which lead to a larger discrepancy). We illustrate this in
Fig.~8, which shows the $A^{FB}_b$ pull vs. the $R_b$ pull in the
supergravity model.

\begin{figure}[htb]
\vbox{\kern4cm\includegraphics{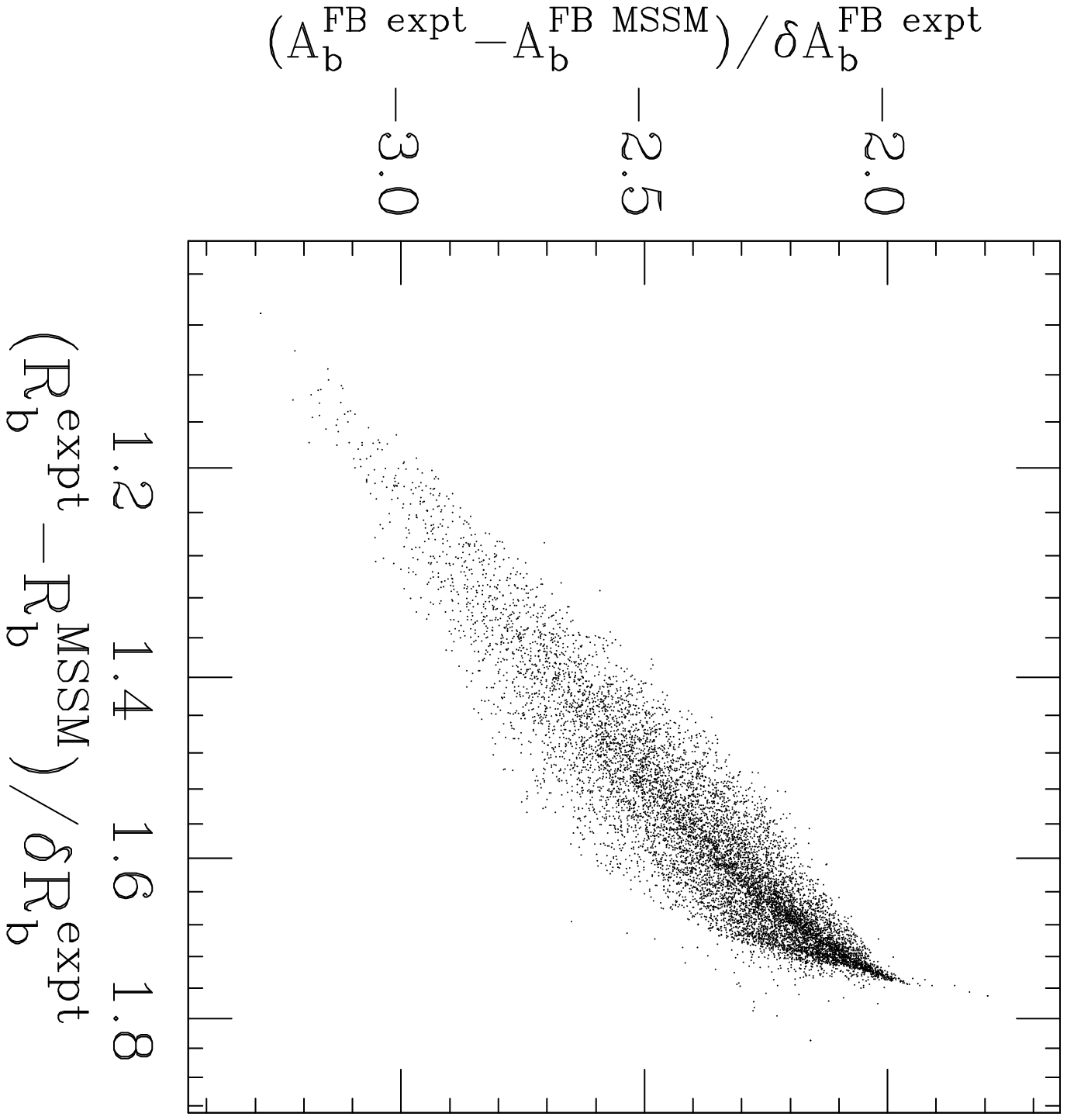}}
\caption{The pull of $A^{FB}_b$ vs. the pull of $R_b$ in the
supergravity model.}
\end{figure}

The minimum $\chi^2$ found in our random scan over parameter space is
29.6 for 19 degrees of freedom in the supergravity model. For both GM
models, we find the minimum $\chi^2$ values of 30.3 for 20 degrees of
freedom. These correspond to goodness of fits of 5.8\% and 6.5\%,
respectively. We can compare these numbers with the standard model
goodness of fit of 10.9\%. We see that the marginal reduction
in $\chi^2$ is more than compensated for by the increase in the number
of input parameters.

\begin{figure}[htb]
\vbox{\kern5.2cm\includegraphics{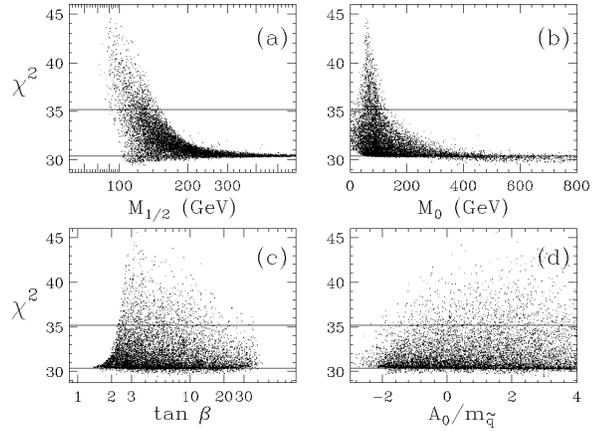}}
\caption{The total $\chi^2$ vs. the input parameters in the
supergravity model with $\mu>0$. The lower horizontal line shows the
minimum $\chi^2$ in the standard model, with the external constraint,
$\alpha_s = 0.118 \pm 0.003$, included. The points above the upper
horizontal line are ruled out at the 95\% CL.}
\end{figure}

\begin{figure}[htb]
\vbox{\kern5.2cm\includegraphics{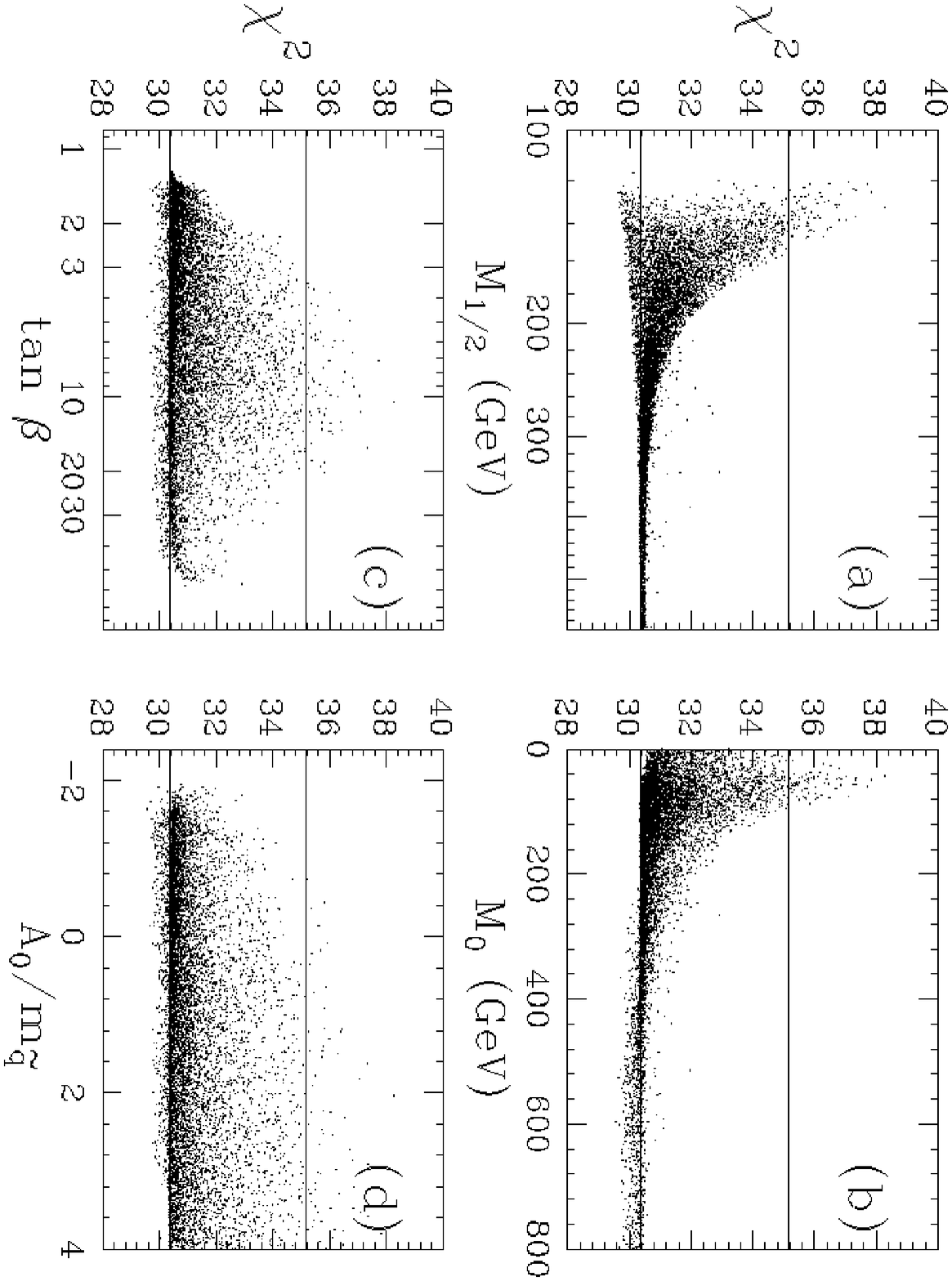}}
\caption{The same as Fig.~9, with $\mu<0$.}
\end{figure}

What we are really interested in is the set of points with large
values of $\chi^2$. Not counting the supersymmetry parameters as fit
parameters, we deem a point in supersymmetric parameter space excluded
if the goodness of the fit is less than 5\%. In Fig.~9 we show the
plots of $\chi^2$ vs. the input parameters in the supergravity model,
with $\mu>0$. All points above the upper horizontal line are excluded
at the 95\% confidence level.  Fig.~10 shows the same plot with
$\mu<0$. With $\mu>0$ there is significantly more parameter space
ruled out. We see that in the $\mu>0$ ($\mu<0$) case, there are no
points excluded if $M_{1/2}> 155 \ (160)$ GeV, or $M_0>160\ (100)$
GeV, or $\tan\beta<2.2\ (3)$.  The excluded parameter space forms a
complicated hyper-region.

\begin{figure}[htb]
\vbox{\kern2cm\includegraphics{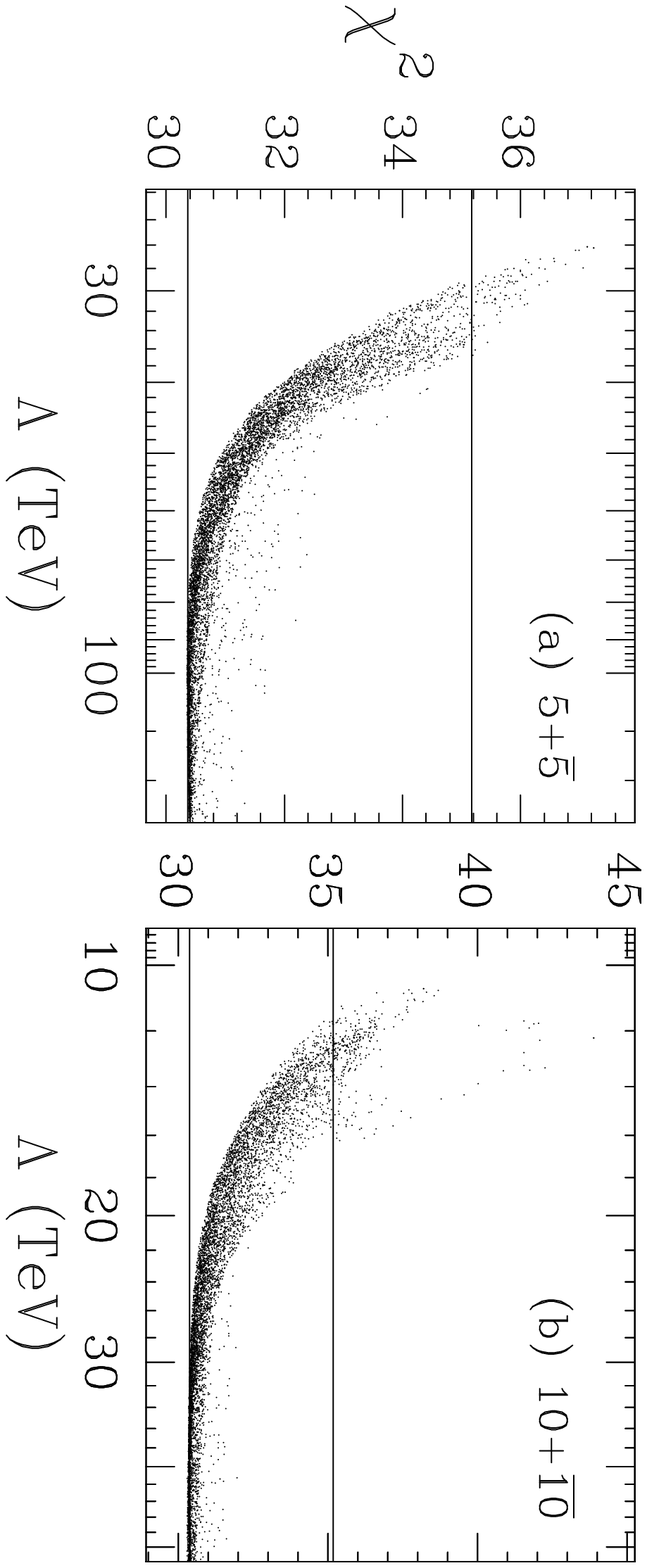}}
\caption{The total $\chi^2$ vs. $\Lambda$ in the (a) \fiv and (b) \ten
gauge-mediated models. The horizontal lines are as in Fig.~9.}
\end{figure}

We plot the full one-loop $\chi^2$ vs. $\Lambda$ in the \fiv and \ten
GM models in Fig.~11, with $\mu>0$. We see that in the \fiv model
values of $\Lambda<30$ TeV are excluded. In the \ten case,
$\Lambda<12$ TeV is excluded. In the \fiv GM model with $\mu<0$ there
are no points excluded by this analysis.

\section{The $B\rightarrow X_s\gamma$ constraint}

\begin{figure}[htb]
\vbox{\kern2cm\includegraphics{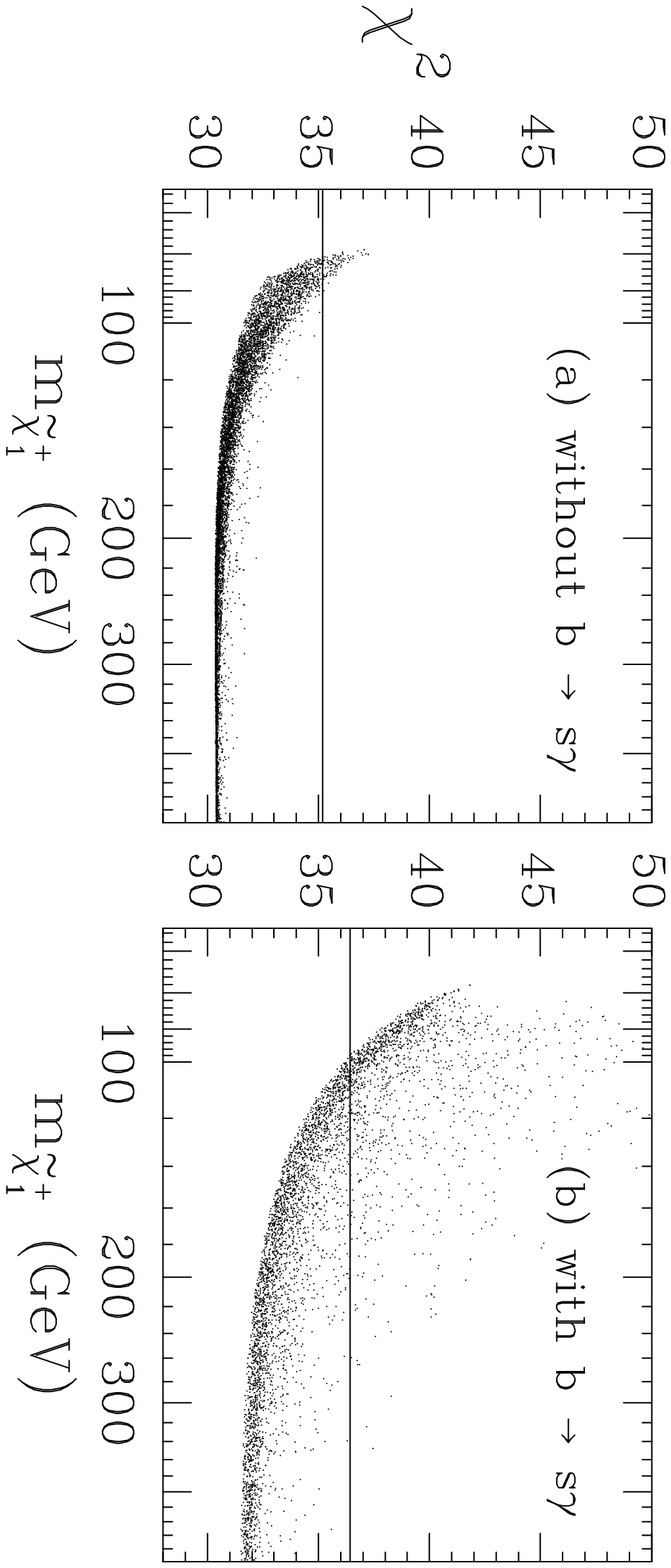}}
\caption{The total $\chi^2$ vs. the light chargino mass in the \fiv
gauge-mediated model, with and without including the $B(B\rightarrow
X_s\gamma)$ constraint. The 95\% CL region lies below the horizontal
line.}
\end{figure}

\begin{figure}[htb]
\vbox{\kern3cm\includegraphics{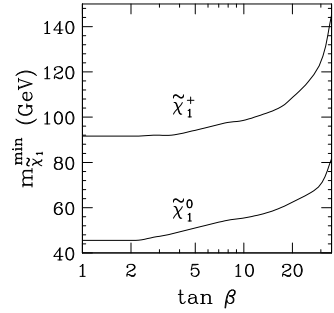}}
\caption{The lower bound on the lightest neutralino mass and the
lightest chargino mass vs. $\tan\beta$, in the gauge-mediated model
with the $B(B\rightarrow X_s\gamma)$ constraint and $\mu>0$.}
\end{figure}

We now specially consider a very important observable.  The CLEO
measurement \cite{DMP:cleo} of the rare decay $B\rightarrow X_s\gamma$
yields the 90\% confidence interval $1.0\times10^{-4} < B(B\rightarrow
X_s\gamma) < 4.2\times10^{-4}$.  This measurement imposes a
significant constraint on supersymmetric models \cite{DMP:bsg}.  The
charged Higgs loops and chargino loops give the largest
contributions. The chargino contribution contains a term proportional
to $\mu\tan\beta$, and it can be much larger than the standard model
amplitude, and can be of either sign. Hence, the $B\rightarrow
X_s\gamma$ rate in supersymmetric models can be much larger or much
smaller than the standard model prediction. This leads to very large
values of $\chi^2$ in some regions of parameter space (e.g. $\mu>0$
and large $\tan\beta$). This is illustrated in Fig.~12, where we show
the $\chi^2$ before and after considering the $b\rightarrow s\gamma$
constraint, in the \fiv GM model, with $\mu>0$. We see from Fig.~12(b)
that we can exclude chargino masses below $M_Z$ with the $b\rightarrow
s\gamma$ constraint. For larger values of $\tan\beta$, this constraint
becomes stronger. We show in Fig.~13 the lower bounds on the lightest
neutralino and lightest chargino masses vs. $\tan\beta$. There are
corresponding bounds on the other superpartner masses. There are no
such bounds in the case $\mu<0$.

In the supergravity model the $b\rightarrow s\gamma$ constraint does
not yield such strong limits, because the parameter space has more
freedom. Nevertheless, the additional region of parameter space which
is excluded after including the $b\rightarrow s\gamma$ measurement is
significant. In particular, the $\mu>0$, large $\tan\beta$ region is
severely constrained.

\section{Conclusions}

Global fits to the world's precision data provide significant
constraints on supersymmetric models. We gave an encapsulated view of
the supersymmetric corrections by examining the oblique set. We then
indicated the amount of parameter space which is excluded based on a
full one-loop analysis. We found it important to include as many
observables as possible in the fit, since different models, or
different regions of parameter space in a given model, are more or
less sensitive to different observables.

We showed the added sensitivity after including the $b\rightarrow
s\gamma$ measurement in the list of observables. The large
$\tan\beta$, $\mu>0$ region of parameter space was shown to be
severely constrained.

Because the supersymmetric corrections decouple and the standard model
with a light Higgs boson is consistent with the data, most of the
supersymmetric parameter space is consistent with the data.  Some
regions of parameter space with light superpartners are excluded, but,
on the other hand, there are points in parameter space with very light
particles which are consistent with the data. In fact, the point in
the supergravity model parameter space with the smallest $\chi^2$ in
our scan includes a light right-handed top-squark (55 GeV). So who
knows what the next experiments will find?

\section*{Acknowledgements}
D.M.P. thanks the CERN theory group and the CERN computing center
staff for generous hospitality.

\end{document}